\documentclass[12pt]{article}

\usepackage{amsmath}
\usepackage{amssymb}
\usepackage{amsfonts}
\usepackage{graphicx}
\usepackage{dcolumn}
\usepackage{bm}

\newcommand{\bk}{\vett k}
\newcommand{\bh}{\vett h}
\newcommand{\bx}{\vett x}
\newcommand{\vett}[1]{\mathbf{#1}} 
\def\beq{\begin{equation}}
\def\eeq{\end{equation}}

\def\eps{ { \varepsilon } }

\def\phi{{\varphi}}
\def\equal{\buildrel {\rm def} \over {=} }
\def\={ \equal } 

\title{ Agreement of classical Kubo theory\\
 with the infrared   dispersion curves $n(\omega)$  of ionic crystals  }

\author{    
 Andrea Carati\thanks{Department of Mathematics, Universit\`a degli
 Studi di  Milano, Via Saldini 50, 20133 Milano -- Italy. Corresponding author: \texttt{andrea.carati@unimi.it}
 }%
 \and Luigi  Galgani\footnotemark[1]    %
 \and Roberto Gangemi\footnotemark[2] 
 \and Alberto Maiocchi\footnotemark[1] %
 \and Fabrizio Gangemi\thanks{DMMT, Universit\`a di Brescia, Viale Europa 11, 25123 Brescia -- Italy} %
}
     
\date{31/03/2015}

\begin{document}

\maketitle




\begin{abstract}
The theoretical  dispersion curves $n(\omega)$ 
(refractive  index  versus frequency) of  ionic crystals in the 
infrared domain are expressed, within  the
Green--Kubo theory,   in terms of a   time
correlation function involving the motion of the ions only.
The aim of this paper is to investigate how well the experimental data are
reproduced by a classical approximation of the theory, in which the time
correlation functions are expressed in terms of the ions orbits.
We report the results of molecular dynamics (MD) simulations for the ions
motions of  a LiF lattice of 4096 ions at room temperature. The
theoretical curves thus obtained  are in  surprisingly good  agreement with 
the experimental data,  essentially  over the whole infrared domain.
This shows that at room temperature the motion of the ions
develops essentially  in a classical regime. 
\end{abstract}

\vfill
\vskip 1em
\noindent
PACS: 63.20.dk, 63.20.Ry, 78.20.Ci

\newpage


Experimental data on the dispersion curves $n(\omega)$ -- refractive
index versus angular frequency -- of ionic crystals in the infrared
domain have been available for a long time \cite{palik}. However, a
microscopic explanation covering the whole infrared domain is
apparently lacking, notwithstanding the fact that the general
theoretical frame is well  defined. On the other hand,  a recent work
 \cite{prl2} (see also \cite{prl1})  has shown that the dispersion
 relation $\omega(\bk)$ in  ferroelectrics can be calculated
 through MD simulations.  In this paper we show that
a calculation of  the ions motions through MD
simulations leads to theoretical dispersion curves $n(\omega)$ for
ionic crystals 
which agree in a surprisingly good way with the experimental data
essentially over the whole infrared domain. 
In fact,  the problem can be dealt with through MD
simulations because   by Green--Kubo theory the refractive index is
expressed  in terms of a time correlation function 
involving  the microscopic dipole moment  $\vett P$ due to the ions motions.

Let us recall the theoretical frame. The refractive index is just the
square root of the electric permittivity tensor
$\eps=\eps_{ij}(\omega)$, which in turn  is related  
to  the  dielectric susceptibility tensor $\chi(\omega)$ through
\begin{equation}\label{perme}
\eps_{ij}(\omega)=\delta_{ij}+4\pi \chi_{ij}(\omega)\ .
\end{equation}
Now, the susceptibility tensor $\chi(\omega)$ is the response function
of the considered medium to an external electric field, so that the
Green-Kubo theory provides for it a formula involving a microscopic
quantity, the polarization $\vett P(t)$.  
For a ionic crystal
the contribution to  polarization in the infrared range
comes mainly from the  ions. So the microscopic polarization can be
written as
\begin{equation}\label{pola}
\vett P= \frac 1{V} \sum_{\bh,s} e_s \vett q_{s,\bh}\ ,
\end{equation}
where $\vett q_{s,\bh}$ is the displacement of the  ion of the
$s$ species (having charge $e_s$) in the  $\bh$--th cell from its
equilibrium position,  and the summation is
understood over all  ions belonging to the (small) volume $V$.

The quantum theoretical formula for the ionic
contribution to  susceptibility  is given  
 for instance  in  \cite{mara} (see formula
(2.16a)) and in \cite{gordon}. For a review see \cite{wallis}.   
The corresponding   
classical limit  is then (see formula (2.16b) in \cite{mara}, or 
see \cite{andrea} for a purely classical deduction) 
\begin{equation}\label{kubo}
\chi_{ij}(\omega) =   \beta V\int_0^{+\infty} e^{-i\omega t}
\langle  P_i(t) \dot P_j(0)\rangle d t \ .
\end{equation}
Here, as usual, $\beta=1/k_BT$ is  inverse temperature 
with $k_B$ the Boltzmann constant, while
$\langle \ldots \rangle$ denotes ensemble average.
\begin{figure}
  \begin{center}
    \includegraphics[width = 0.9\textwidth]{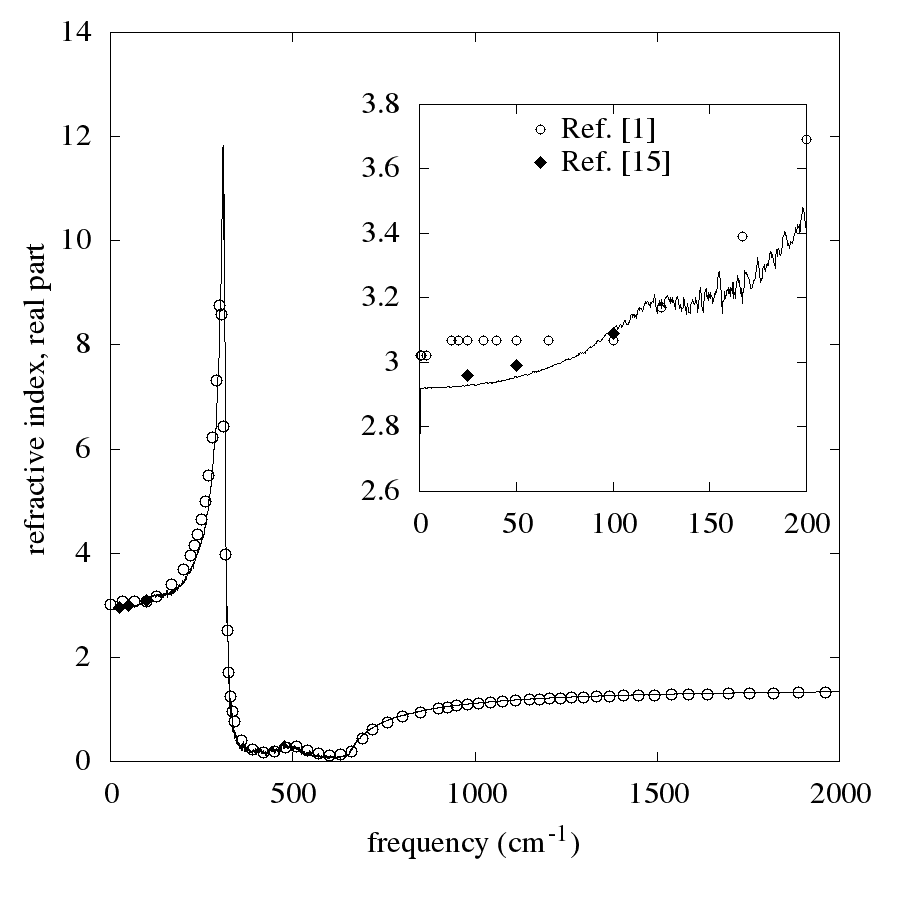}
  \end{center}
  \caption{\label{figura2} Real part of the refractive  index versus
    frequency for a LiF crystal at 300 K.
 Circles and diamonds are experimental data taken from
 \cite{palik}  and \cite{kachare} respectively. Solid line is 
 obtained, through Green--Kubo theory, 
from   MD simulations for the motions of a system of 4096 ions.
The inset is a zoom of the leftmost part of the figure.}
\end{figure}
Instead,  the contribution of  the  electrons to susceptibility  
in  the infrared is well known from experiments to just reduce 
to a constant (isotropic) tensor 
 $\chi^{(\infty)}\delta_{ij}$, which  can be taken from \cite{palik}.   

Thus the computation of the refractive index
in the infrared range is  reduced to a computation of the ionic  
susceptibility, which in turn is reduced, through (\ref{kubo}) and
(\ref{pola}), to a determination of the classical  motions of the ions.

So, we model the crystal as a face centered cubic lattice of $2N$
particles (the ions, with their atomic
masses) in a cubic box of side $L=l_0(N/4)^{\frac 13}$, where $l_0$ is
the lattice constant\footnote{Recall that the \emph{primitive} cell of a
  fcc crystal in not cubic at all but tetrahedral, and that the unit cubic
  cell is arranged using four primitive cells.}. 
The ions interact  through mutual electric
forces, and through a  phenomenological short--range
repulsive pair  potential $V_{rep}$. 
Periodic boundary conditions are imposed,
and the equilibrium positions $\bx_{s,\bh}$ of the atomic species
$s=1,2$ in the cell $\bh\in \mathbb Z^3$ of the lattice, as well as
the value of the lattice constant $l_0$, are taken from the literature
\cite{poseq}. In the numerical simulations, we took 
$2N=4096$.

For the aim of estimating the electric forces between ions,
it is known (see \cite{szigeti}) that the ions can be dealt with as
point particles only if a suitable ``effective charge'' $e_{eff}$ is
substituted for the ions charge.  
In this paper we concentrate on lithium
fluoride, which is the halide presenting the simplest infrared
spectrum, using for it  the  value given in the literature (see
\cite{lowndes}), namely $e_{eff}=0.81 e$, where
 $e$ is  the electron charge.

Concerning the phenomenological  repulsive  short--range pair
potential  $V_{rep}$, in the literature one  often takes it  
to be spherically symmetric, independent of the ions'
species, and  depending on the distance  $r$ as an exponential,
$a\exp (-r/r^*)$. Here we chose the analytic form previously  considered  
by  Born  in his  MIT  lectures \cite{born} and in the paper
\cite{alessio}, namely,   
$V_{rep}(r)=a\,r^{-p}$, with a  cut--off at 5 \AA. Thus, having fixed
the effective charge, the model still
contains  two free parameters,  $a$ and  $p$.
We are confident that the choice of the repulsive potential,
exponential or inverse power, doesn't play an essential role.

The two parameters  $a$ and  $p$ can in principle be
determined through the thermodynamic available data, for example
from the compressibility and the heat of formation.
Here we determined them
from the experimental dispersion relation $\omega(\bk)$ 
obtained from neutron scattering. To this end we made 
reference to the work \cite{alessio}, which  was devoted to an analytical 
estimate of the dispersion relation in the linear approximation with
the aim of  a microscopic explanation of the existence of
polaritons. In that work
  the elastic constants of the crystal had to be determined by fit
with the experimental curves. As the 
elastic constants are related to the first two derivatives of the
phenomenological potential $V_{rep}$
(see \cite{libroincomprensibile}), this information, together 
with the choice that
$p$ be an integer, allowed to determine $a=1.25$ and $p=6$.

\begin{figure}
  \begin{center}
   \includegraphics[width = 0.9\textwidth]{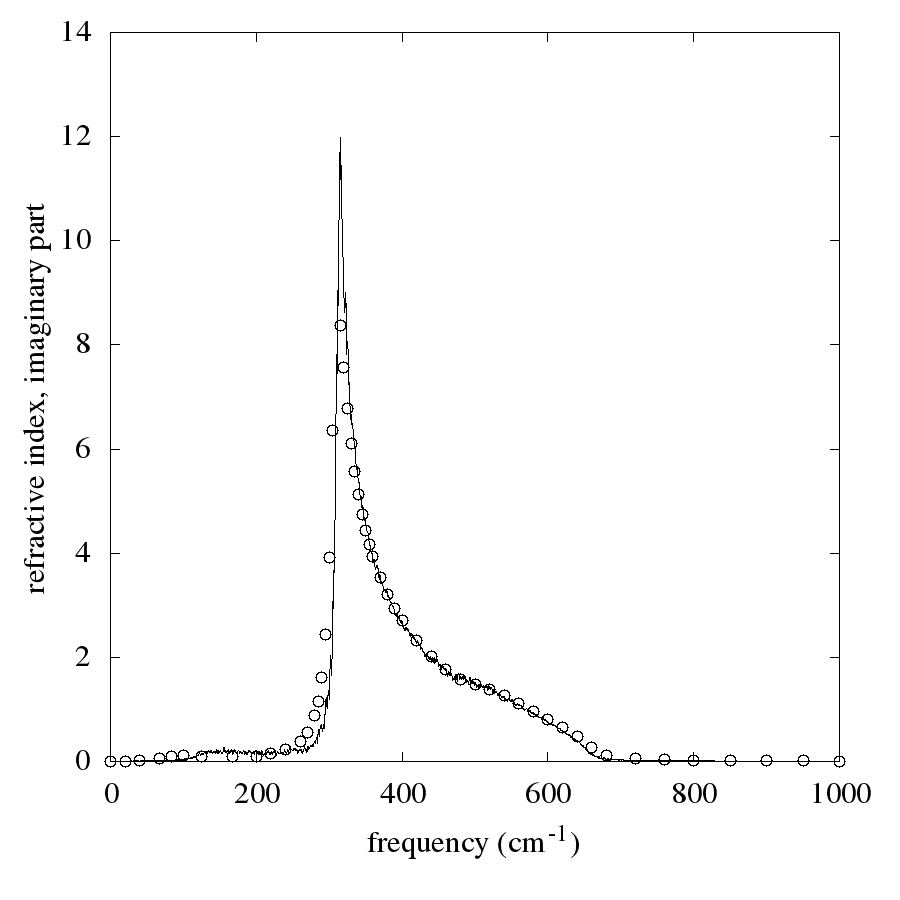}
 \end{center}
  \caption{\label{figura2b} 
Imaginary part of the refractive  index versus
    frequency for a LiF crystal at 300 K. As in figure \ref{figura2},
 circles are experimental data taken from
 \cite{palik}, while  solid line is 
 obtained, through Green--Kubo theory, 
from   MD simulations for the motions of a system of 4096 ions. 
}
\end{figure}
\begin{figure}
  \begin{center}
    \includegraphics[width = 0.9\textwidth]{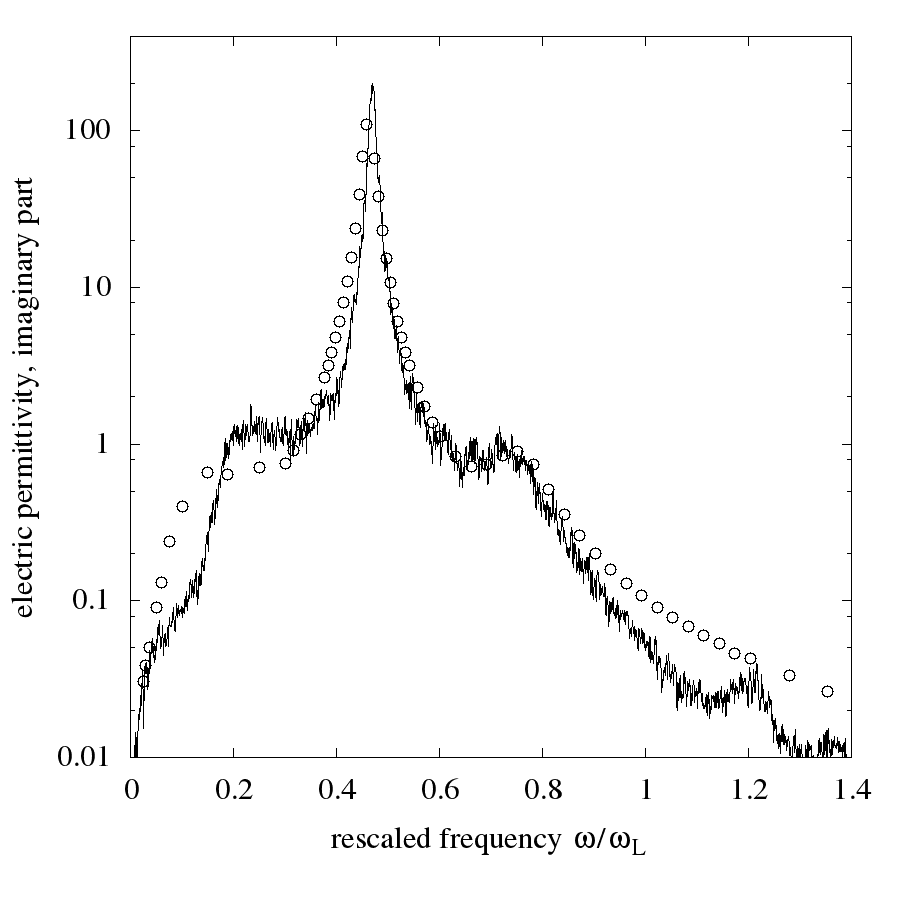}
  \end{center}
  \caption{\label{figura3} Imaginary part of the electric permittivity
 $\eps$ (in logarithmic scale) versus rescaled frequency $X=\omega/\omega_L$, where
    $\omega_L=665$ cm$^{-1}$  
is the frequency of the zero longitudinal optical mode. Circles are
experimental data from \cite{palik}, while solid line is 
 obtained, through Green--Kubo theory, 
from   MD simulations for the motions of a system of 4096 ions.
Compare with the theoretical curve reported in  Fig.~12 of 
 ref.~\cite{mara2}, obtained with quantum mechanical methods. Note
that the experimental data reported there  are the older ones  
\cite{gott} and \cite{genzel}. The agreement with the new data
 \cite{palik} reported here is apparently  better.}
\end{figure}

Concerning the electric  forces, due 
to their long--range nature  a problem arises 
in writing down explicitly the equations of motion  when  
periodic boundary conditions are imposed, because 
one then  has  to determine  the electric field produced by an
infinite lattice of charges.  As is
well known  (see \cite{ewaldnumerico}), the problem is solved in a
computationally efficient way using   the  Ewald resummation
formula. This amounts to introducing  the potential
\begin{equation*}
\begin{split}
V_{Coul}=\sum_{s,\bh}\sum_{s',\bh'}(-1)^{s+s'}e^2_{eff}\left[
\sideset{}{'}\sum_{\vett  n\in \mathbb Z^3} 
\frac{\mathrm{erfc}(\delta
  |\vett r^{(s,s')}_{\bh,\bh'}
  +\vett n\, L |)}{|\vett r^{(s,s')}_{\bh,\bh'}  +\vett n\, L |} \right.\\
+ \left.\frac{4\pi}{L^3} \sum_{\bk\in \mathbb Z^3}
 \exp\left(-\frac{\pi^2|\bk|^2}{\delta^2L^2}
 \right)\exp\left(i\frac{2\pi}{L}\bk\cdot\vett r^{(s,s')}_{\bh,\bh'}
 \right) \right]
 \end{split}
\end{equation*}
where $\mathrm{erfc}(x)$ is the complementary error function, $\delta$
is a parameter to be chosen to insure fast convergences of the series,
$\vett r^{(s,s')}_{\bh,\bh'} =(\vett x_{s,\bh}+\vett q_{s,\bh})-(\vett
x_{s',\bh'}+\vett q_{s',\bh'})$ and the primed sum over 
$\vett n$ means that the term for $\vett n=0$ is omitted if $s=s'$ and
$\bh=\bh'$. We took $\delta=0.2$ \AA$^{-1}$ which allows to truncate the
series at $|n_x|,|n_y|,|n_z|\le N_{max}=1$ and $|k_x|,|k_y|,|k_z| \le
K_{max}=3$.

The numerical integrations of the equations of motion were performed
with a  standard symplectic Verlet method. The integration step was
taken equal to $2$ fs, the duration of each simulation was  of 100 ps,
and the time averages (which will be mentioned in a moment) were taken
over 50 ps.  The initial data for 
each orbit
were assigned by setting the particles in their equilibrium positions,
while the velocities were extracted according to the Maxwell--Boltzmann
distribution, with
the constraint that the center of mass of the system has vanishing
velocity. Then, the temperature of the system (300 K) 
was  determined  through the
mean kinetic energy,  once  thermalization had been attained.

So we computed the orbits, i.e.  the displacement $\vett q_{s,\bh}(t)$  of
the ions from their  equilibrium positions, for a
LiF lattice of 4096 ions at 300 K, and
thus the microscopic polarization $\vett P(t)$ too,  given by (\ref{pola}). 
In formula \eqref{kubo} for the susceptibility the ensemble  average 
was estimated as the mean value, over 10 different orbits, of the
corresponding time average.   

As a check for our computations we controlled that the
susceptibility tensor  actually is, as expected for LiF, an isotropic one.
Namely,  the off--diagonal matrix elements are negligible and the
diagonal ones can be considered as equal.  
One may presume that
these properties should not occur in a less symmetric type of crystal,
so that phenomena of optical activity  would  show up in such cases.
In our case, however,  the susceptibility
tensor can be dealt with as a scalar, which was estimated  as 
the mean of the three diagonal  matrix elements. 

The results of the computations are now illustrated. 
In figure~\ref{figura2} we show the main result of
this paper, namely, the real part   of
the refractive index $n=\sqrt\epsilon$, with $\epsilon$ given by  
(\ref{perme}) and $\chi$ computed through (\ref{kubo}). The
theoretical curve was plotted  together
with the experimental data taken from \cite{palik} and from \cite{kachare}.  
The error bars of
the experimental data are not reported because 
 the relative errors are  negligible (less
than 0.1 \%). Notice that, in the theoretical curve, only the
parameter $\chi^{(\infty)}$ was directly fitted to the set of
experimental data, taking it from the data reported in 
\cite{palik}.

The overall agreement  appears to be 
surprisingly good, essentially  over the whole infrared domain.  
Actually one may notice that, as exhibited by the inset,  
apparently the agreement is not
that  good in the far infrared range, particularly  below 100
cm$^{-1}$. In this
connection we point out first of all
that  in such a region  the experimental data  appear 
not to be so sure
because in the literature different
series of measures are found, which are not mutually compatible in
view of  the
declared errors. Our theoretical curve are however in fairly good
agreement with the experimental data of reference \cite{kachare},
which are, at any rate, the more recent ones.

Should the measurements of ref.~\cite{kachare} be the good ones, then
a very interesting perspective would be opened. Indeed it is known
that   for very low frequencies (below 1\,MHz) the refractive index  is
slightly above 3 (between 3.05 and 3.018 according to 
\cite{statico}), while in the far infrared range we found, in agreement with
ref.~\cite{kachare},   a value near 2.9, i.e. below the
static limit $\omega\to 0$. So, the refractive index should increase
as $\omega$ tends to zero, and this indeed happens in the microwave
range (GHz), according to \cite{jacob}, where a value slightly below 3 is
found.

To settle this question, at least at the level of MD simulations, integrations
of the equations of motion over much longer time scales would be
necessary. By the way, the problem of the dependence of the results on the
integration time  is a very delicate one,
and was discussed, for example, in \cite{sascha} in connection with a MD
computation of  the specific heat of an  Argon crystal, where  a comparison
between MD simulations and experimental data was performed somehow in the same
spirit of the present paper. Another possibility to tackle the problem
 is to enlarge the model,  in order to take some further
physical property into account. We are thinking of the
retarded character of the electromagnetic forces. In fact, in the work
\cite{alessio} retardation turned out to be the essential qualitative
ingredient in proving the existence of polaritons (i.e., the presence
of two new branches in the dispersion relation; see \cite{am} and
\cite{pastori}) in a microscopic model. We are analogously suggesting
that retardation might lead to the arising of new absorption bands in
the far infrared range. We hope to come back to these  problems  in the
future.

In any case, leaving aside  the problems related to  the   far
infrared, the present  results  show that the classical
 motions of the ions at room temperature, estimated by MD simulations
 on  a sample of 4096 ions at 300 K, lead to  theoretical
dispersion curves which are, overall,  in very good  agreement with the
experimental data. We  recall that theoretical estimates of the
 dispersion curves of ionic crystals   were given also in the
frame of quantum mechanics for  the same model.  
 To our knowledge  such theoretical estimates,
which use many-phonon perturbation methods, presently do
not cover the whole infrared domain. In the domain
in which they  apply, the results are apparently not  better 
than the present ones.  This is illustrated through figure~\ref{figura3},
which should be compared with the quantum theoretical predictions for
LiF at room temperaturereported in Fig. 12 of \cite{mara2}.

\vskip 1em
\noindent
\textbf{Acknowledgements}: we thank G. Grosso, G. Pastori Parravicini and N. Manini for
  useful discussions. The use of computing resources provided by the
  Italian Grid Infrastructure (IGI) is also  gratefully acknowledged.


\end{document}